\documentclass{aastex631}

\usepackage{chemformula}

\newcommand\mps{m~s$^{-1}$}

\newcommand\kgpmmm{kg~m$^{-3}$}

\newcommand\coo{\mbox{CO$_2$}}

\newcommand\rh{\ensuremath{r_{\mathrm{h}}}}
\newcommand\inv[2][1]{$\textrm{#2}^{-#1}$}
\newcommand\timesten[1]{\mbox{$\times10^{#1}$}}
\newcommand\specgrad[2]{$S_{#1}=#2$\% per 100 nm}

\newcommand\BB{Bernardinelli-Bernstein}

\accepted{June 23, 2022}
\submitjournal{ApJL}

\shorttitle{Outbursts of C/2014 UN$_{271}$ (Bernardinelli-Bernstein)}
\shortauthors{Kelley et al.}
\graphicspath{{./}{figures/}}

\begin{document}

\title{A LOOK at Outbursts of Comet C/2014 UN$_{271}$ (Bernardinelli-Bernstein) Near 20~au}

\author[0000-0002-6702-7676]{Michael S. P. Kelley}
\affil{Department of Astronomy, University of Maryland, College Park, MD 20742-0001, USA}
\email{msk@astro.umd.edu}

\author[0000-0003-4617-8878]{Rosita Kokotanekova}
\affiliation{European Southern Observatory, Karl-Schwarzschild-Str 2, 85748 Garching, Germany}
\affil{Institute of Astronomy and National Astronomical Observatory, Bulgarian Academy of Sciences, 72 Tsarigradsko shose Blvd., 1784 Sofia, Bulgaria}

\author[0000-0002-4043-6445]{Carrie E. Holt}
\affil{Department of Astronomy, University of Maryland, College Park, MD 20742-0001, USA}

\author[0000-0001-8541-8550]{Silvia Protopapa}
\affiliation{Southwest Research Institute, 1050 Walnut Street, Suite 300, Boulder, CO 80302, USA}

\author[0000-0002-2668-7248]{Dennis Bodewits}
\affiliation{Physics Department, Leach Science Center, Auburn University, Auburn, AL 36849, USA}

\author[0000-0003-2781-6897]{Matthew M. Knight}
\affiliation{Physics Department, United States Naval Academy, 572C Holloway Rd, Annapolis, MD 21402, USA}

\author[0000-0002-3818-7769]{Tim Lister}
\affil{Las Cumbres Observatory, 6740 Cortona Drive Suite 102, Goleta, CA 93117, USA}

\author[0000-0002-8658-5534]{Helen Usher}
\affiliation{The Open University, Walton Hall, Milton Keynes, MK7 6AA, UK}

\author[0000-0002-1278-5998]{Joseph Chatelain}
\affil{Las Cumbres Observatory, 6740 Cortona Drive Suite 102, Goleta, CA 93117, USA}

\author[0000-0001-5749-1507]{Edward Gomez}
\affil{Las Cumbres Observatory, School of Physics and Astronomy, Cardiff University, Queens Buildings, The Parade, Cardiff CF24 3AA, UK}

\author[0000-0002-4439-1539 ]{Sarah Greenstreet}
\affil{Department of Astronomy and the DIRAC Institute, University of Washington, 3910 15th Ave NE, Seattle, WA 98195, USA}

\author[0000-0003-4881-6255 ]{Tony Angel}
\affiliation{Harlingten Observatory, Observatorio Sierra Contraviesa, Cortijo El Cerezo, Torvizcon 18430, Granada, Spain}

\author{Ben Wooding}
\affiliation{St Mary's Catholic Primary School, Llangewydd Road, Bridgend, Wales, CF31 4JW, UK}

\begin{abstract}
  Cometary activity may be driven by ices with very low sublimation temperatures, such as carbon monoxide ice, which can sublimate at distances well beyond 20~au.  This point is emphasized by the discovery of Oort cloud comet C/2014 UN$_{271}$ (Bernardinelli-Bernstein), and its observed activity out to $\sim$26~au.  Through observations of this comet's optical brightness and behavior, we can potentially discern the drivers of activity in the outer solar system.  We present a study of the activity of comet \BB{} with broad-band optical photometry taken at 19--20~au from the Sun (2021 June to 2022 February) as part of the LCO Outbursting Objects Key (LOOK) Project.  Our analysis shows that the comet's optical brightness during this period was initially dominated by cometary outbursts, stochastic events that ejected $\sim10^7$ to $\sim10^8$~kg of material on short ($<1$~day) timescales.  We present evidence for three such outbursts occurring in 2021 June and September.  The nominal nuclear volumes excavated by these events are similar to the 10--100~m pit-shaped voids on the surfaces of short-period comet nuclei, as imaged by spacecraft.  Two out of three Oort cloud comets observed at large pre-perihelion distances exhibit outburst behavior near 20~au, suggesting such events may be common in this population.  In addition, quiescent CO-driven activity may account for the brightness of the comet in 2022 January to February, but that variations in the cometary active area (i.e., the amount of sublimating ice) with heliocentric distance are also possible.
\end{abstract}


\section{Introduction} \label{sec:intro}

Comet C/2014 UN$_{271}$ (Bernardinelli-Bernstein) was discovered in a search of the Dark Energy Survey data archive for new Solar System objects \citep{bernardinelli21-c2014un271}.  Originally designated as an asteroid, the object was found in data from  2014 to 2018, at heliocentric distances, \rh, of 29.0 to 23.7~au.  Shortly after the discovery announcement in 2021 \citep{mpec2021-M53}, new observations at \rh=20~au found the object to be extended and about --1.5~mag brighter than expected based on the 2014--2018 photometry, and the object was subsequently re-designated as a comet \citep{kokotanekova21-atel14733,buzzi21-cbet4989}.  The object was serendipitously observed by the Transiting Exoplanet Survey Satellite (TESS) in 2018 and 2020, showing direct evidence for activity as far out as 23~au, and a lack of any periodic variability \citep{ridden-harper21-un271,farnham21-c2014un271,bernardinelli21-c2014un271}.  A re-analysis of the pre-discovery data in the Dark Energy Survey and Panoramic Survey Telescope and Rapid Response System (PanSTARRS) archives confirmed activity out to $\sim26$~au \citep{bernardinelli21-c2014un271}.  The comet is notable for being discovered at such a great distance \citep[cf.][]{meech17-c2017k2} and for its large size, estimated to be 69$\pm$9~km in radius by \citet{lellouch22-c2014un271}, and 60$\pm$7 to 69$\pm$8~km by \citet{Hui2022}.  It is the largest known Oort cloud comet nucleus, and second to 95P/Chiron \citep{ruprecht15} for the largest cometary nucleus overall.

Little is known about cometary activity at $\sim$20~au.  At such great distances, activity is likely driven by the sublimation of volatiles with low sublimation temperatures.  The volatiles CO and \ch{CO2}, found in abundance in the cometary population \citep{ahearn12-origins}, have nominal sublimation temperatures of 25 and 80~K, respectively, although some level of sublimation can occur below these values \citep{meech04-activity}.   The temperature of a blackbody sphere is $T{\sim}60$~K at 20~au, which suggests CO sublimation can persist out to many tens of au, as long as ices are available near the nuclear surface.  As an alternative to ice sublimation, water ice phase transitions may also drive cometary activity, but they are unlikely to be major contributors at such great distances \citep{jewitt17-c2017k2}.  Aside from comet \BB{}, the three comets with the most distantly observed activities are C/1995 O1 (Hale-Bopp), active post-perihelion out to 27~au \citep{kramer14}, C/2010 U3 (Boattini), active pre-perihelion at 26~au \citep{hui19-c2010u3}, and C/2017 K2 (PanSTARRS), active pre-perihelion at 24~au \citep{hui18-c2017k2}.

We present images and photometry of comet \BB{} taken as part of the LCO Outbursting Objects Key Project \citep[LOOK Project;][]{lister22-look} using the Las Cumbres Observatory (LCO) global telescope network.  We characterize the coma color, lightcurve, and morphology, and show that the comet's activity near 20~au has varied substantially with time.

\section{Observations and Reduction}\label{sec:obs}

Images of comet \BB{} were taken by our team and by school children in Wales as part of the Comet Chasers educational outreach program through the Faulkes Telescope Project.  Observations were taken with LCO's 1-m robotic telescopes, specifically those located at: Siding Spring Observatory, Australia; South African Astronomical Observatory, South Africa; and the Cerro Tololo Inter-American Observatory, Chile.  Each telescope is equipped with identical Sinistro cameras, utilizing 4k$\times$4k CCDs with a nominal pixel scale of 0\farcs39, and a suite of filters.  For most observations, we used the SDSS $g'$ and $r'$ filters, although $w$ (effectively $g+r+i$), $V$, and $R$ filters were used in a few circumstances.  LOOK Project observing sequences were scheduled with the NEOExchange observation manager \citep{lister21-neox}.  The telescopes tracked the proper motion of the comet, and exposure times were typically 300s with 2 exposures per filter.

\begin{figure*}
  \plotone{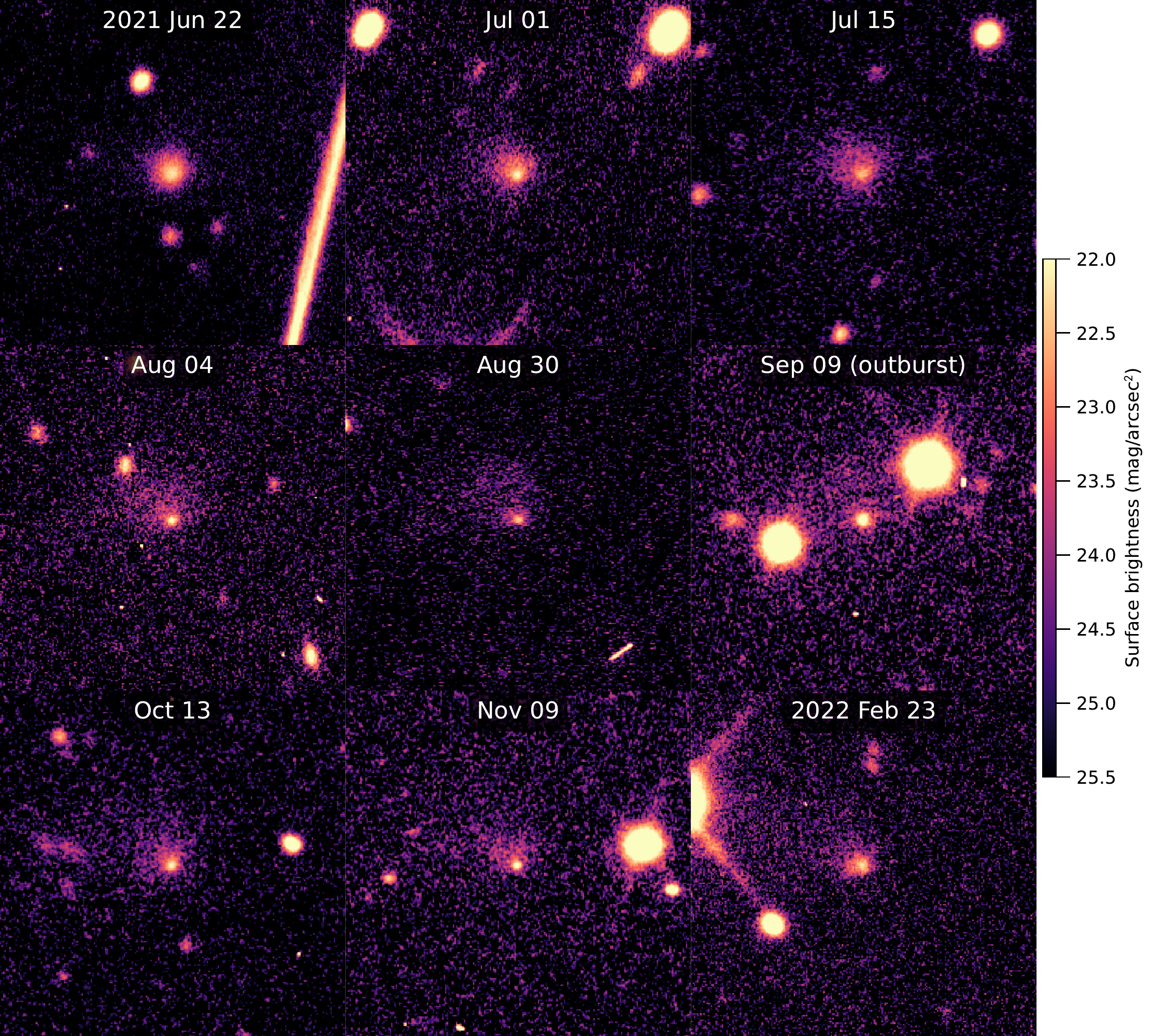}
  \caption{Select LCO images of comet C/2014 UN$_{271}$ (\BB).  Dates are as indicated, all images are through an $r'$ filter, except on 2021 June 22, which is $w$.  All images are displayed with the same color scale and are 1,320,000 km $\times$ 1,320,000 km in size (90\arcsec{} on 2021 June 22).  Celestial north is up, east to the left.  Over this time period, the heliocentric distance decreased from 21.78 to 19.28~au, the projected comet-Sun vector increased from position angle 46\degr{} to 292\degr{}, and the projected velocity vector increased from 144\degr{} to 153\degr{}.}
  \label{fig:images}
\end{figure*}

Images were processed by LCO's BANZAI (Beautiful Algorithms to Normalize Zillions of Astronomical Images) pipeline \citep{McCully2018BANZAI}, which includes bias, dark, and flat-field corrections, source extraction, and astrometric calibration.  Individual frame exposure times were short enough that stars only trail by $\sim0.66$--$0.71\arcsec$ in 300\,s exposures.  We used the pipeline-produced source catalog to photometrically correct each image, calibrating them to the PanSTARRS~1 (PS1) photometric system \citep{tonry12-ps1} using the ATLAS-RefCat2 photometric catalog \citep{tonry18-refcat2}, and LOOK Project derived color corrections.  Details on the data reduction are presented by \citet{lister22-look}.  For this paper, we adopt the apparent magnitude and effective wavelengths for the solar spectrum from \citet{willmer18-sun}: $g_{P1}=-26.54$ AB mag at 481.1~nm, and $r_{P1}=-26.93$ AB mag at 615.6~nm (PS1 system photometry is hereafter denoted $g$ and $r$).  Sample images are presented in Fig.~\ref{fig:images}.

Photometry of the comet was measured in co-added frames, grouped by telescope and time.  We chose an aperture size of 6\farcs0 (86,000 to 88,000~km at the comet; background annulus 20--40\arcsec), and limited our analysis to images with point source FWHMs $<$3\arcsec{} and final photometric uncertainties $<0.15$~mag.  We have 179 photometric measurements in 96 distinct observational epochs.  Our first observation was on 2021~June~22 with the comet at a heliocentric distance \rh=20.18~au, a geocentric distance $\Delta$=20.20~au, and a Sun-target-observer (phase) angle $\theta$=2.9\degr.  Our last observation presented in this paper was on 2022~February~23 at \rh=19.28~au, $\Delta$=19.70~au, and $\theta$=2.6\degr.  The observational circumstances and photometry are presented in Table~\ref{tab:obs}.

\begin{deluxetable*}{cccccccccccccc}
  \caption{Observational circumstances and photometry. \label{tab:obs}}
  \tablecolumns{14}
  \tabletypesize{\footnotesize}
  \colnumbers
  \tablehead{
    Date
    & Year
    & Filter
    & Cal.
    & \rh{}
    & $\Delta$
    & $\theta$
    & Airmass
    & Seeing
    & $N_{\mathrm{images}}$
    & $t_{\mathrm{exp}}$
    & $m$
    & $\sigma_m$
    & $H_r$ \\
    (UTC)
    & (UTC)
    &
    &
    & (au)
    & (au)
    & (\degr)
    &
    & (\arcsec)
    &
    & (s)
    & (mag)
    & (mag)
    & (mag)
  }
  \startdata
  2021-06-22 04:13 & 2021.472 & $w$        & $r_{\mathrm{P1}}$ & 20.177 & 20.198 & 2.884 & 1.25 & 2.7 & 3 & 900 & 18.275 & 0.068 & 5.098 \\
  2021-09-08 02:33 & 2021.685 & $r^\prime$ & $r_{\mathrm{P1}}$ & 19.892 & 19.448 & 2.636 & 1.09 & 1.6 & 1 & 300 & 19.239 & 0.102 & 6.185 \\
  2021-09-09 22:04 & 2021.690 & $g^\prime$ & $g_{\mathrm{P1}}$ & 19.886 & 19.438 & 2.631 & 1.45 & 1.9 & 2 & 600 & 19.090 & 0.072 & 5.564 \\
  2021-09-18 02:09 & 2021.713 & $r^\prime$ & $r_{\mathrm{P1}}$ & 19.856 & 19.401 & 2.617 & 1.10 & 2.2 & 2 & 600 & 18.786 & 0.075 & 5.742 \\
  2021-09-23 05:03 & 2021.727 & $r^\prime$ & $r_{\mathrm{P1}}$ & 19.837 & 19.382 & 2.615 & 1.21 & 1.8 & 2 & 600 & 18.645 & 0.070 & 5.606 \\
  2021-10-07 21:40 & 2021.767 & $r^\prime$ & $r_{\mathrm{P1}}$ & 19.783 & 19.349 & 2.637 & 1.21 & 1.7 & 2 & 600 & 18.883 & 0.065 & 5.852 \\
  2021-11-05 19:45 & 2021.846 & $r^\prime$ & $r_{\mathrm{P1}}$ & 19.678 & 19.367 & 2.764 & 1.21 & 2.0 & 1 & 300 & 19.086 & 0.096 & 6.059 \\
  2021-12-10 19:21 & 2021.942 & $r^\prime$ & $r_{\mathrm{P1}}$ & 19.550 & 19.491 & 2.886 & 1.10 & 2.1 & 2 & 600 & 19.135 & 0.073 & 6.103 \\
  2022-01-11 12:57 & 2022.029 & $r^\prime$ & $r_{\mathrm{P1}}$ & 19.434 & 19.624 & 2.832 & 1.52 & 3.0 & 2 & 600 & 19.184 & 0.080 & 6.153 \\
  2022-02-14 01:28 & 2022.121 & $r^\prime$ & $r_{\mathrm{P1}}$ & 19.312 & 19.700 & 2.667 & 1.55 & 2.6 & 2 & 600 & 18.948 & 0.082 & 5.929 \\
  \multicolumn{14}{c}{\ldots}\\
  \enddata
  \tablecomments{Table 1 is published in its entirety in the machine-readable format.
    A portion is shown here for guidance regarding its form and content.  Column descriptions: (1) mid-time of the observation; (2) mid-time as fractional year; (3) filter used in the observation; (4) PS1 filter to which the photometry is calibrated; (5) heliocentric distance; (6) observer-target distance; (7) Sun-target-observer (phase) angle; (8) calculated airmass of the observation; (9) FWHM of point sources; (10) number of images; (11) total exposure time; (12) apparent magnitude; (13) uncertainty on the apparent magnitude; (14) absolute magnitude in the $r$-band ($g$-band photometry has been scaled by the average color).}
\end{deluxetable*}

\section{Results and Analysis} \label{sec:results}
\subsection{Color}
Based on 70 photometry sets with contemporaneously obtained $g$- and $r$-band photometry, we find a weighted mean coma color of $g-r$=0.47$\pm$0.01~mag.  All color sets are consistent with the mean at the 1.7$\sigma$ level.  There is no evidence for a trend with time or brightness (Pearson and Spearman correlation test p-values $\gg$0.05).  The color of the coma corresponds to a spectral gradient \citep{ahearn84-bowell} of \specgrad{g,r}{5.48\pm0.05}, where subscripts denote the PS1 bandpasses upon which the calculation is based.

\begin{figure*}
  \plotone{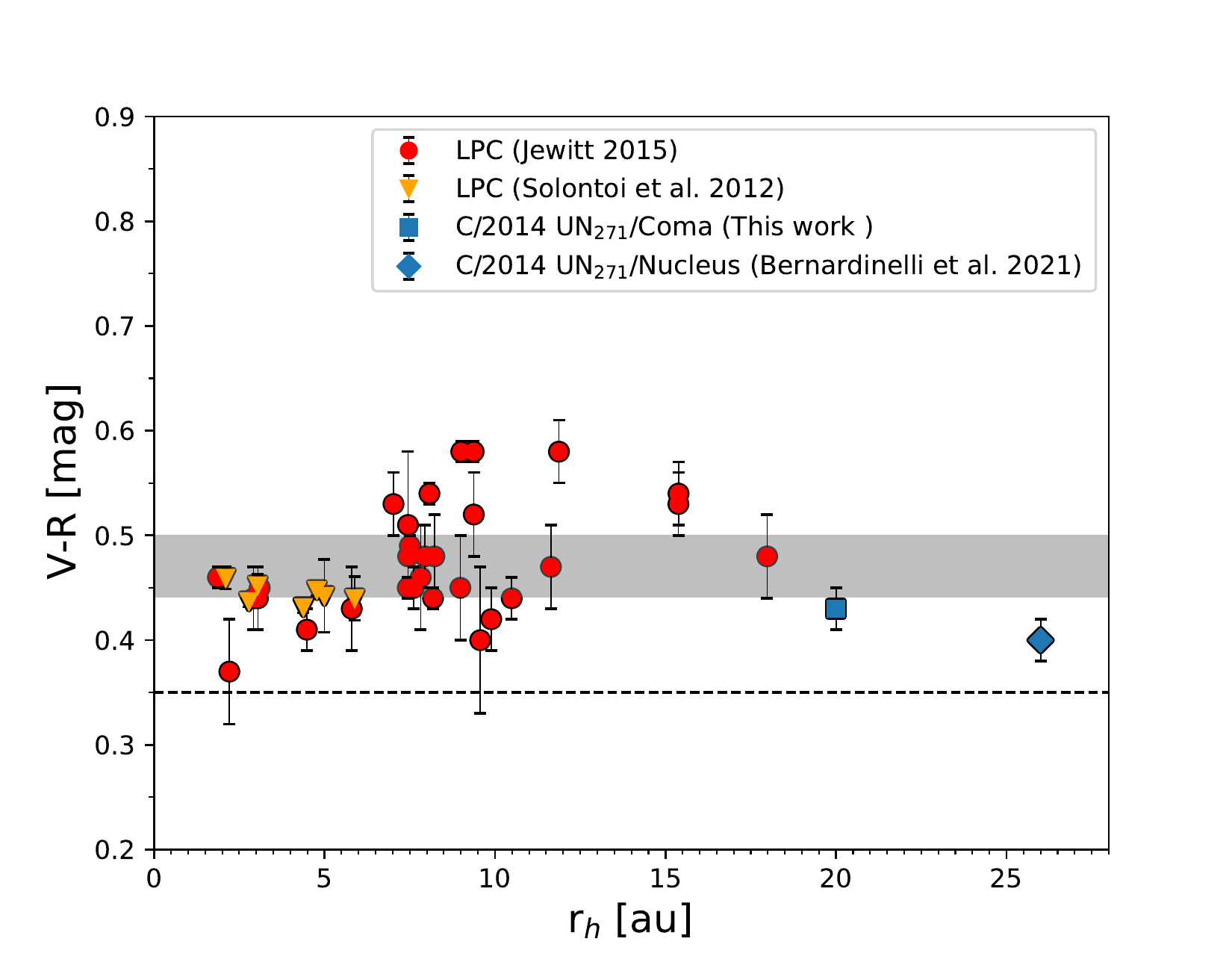}
  \caption{$V-R$ color as a function of heliocentric distance, \rh, comparing the color of the coma of comet C/2014 UN$_{271}$ (\BB) as measured in this work (square symbol) and that of other active LPCs from \citet{jewitt15-color} and \citet{solontoi12-sdss}.  The grey rectangle defines the 3$\sigma$ limits around the average $V-R$ color of active LPCs (0.47$\pm$0.01~mag), and the horizontal dashed line is the color of the Sun.  Color wise, comet \BB{} is a typical active LPC. The color of the nucleus \citep{bernardinelli21-c2014un271} is also shown for comparison (diamond).}
  \label{fig:color}
\end{figure*}

The comparison between the coma color of comet \BB{} and those of other active long period comets \citep[LPCs;][]{solontoi12-sdss,jewitt15-color}, is shown in Fig.~\ref{fig:color}. The comparison is made in terms of $V-R$ color for convenience only.  Using the relations reported by \citet{tonry12-ps1}, the $g-r$ color of comet \BB{} corresponds to $V-R=0.43\pm0.02$~mag, which agrees within 2$\sigma$ of the mean color of active LPCs: $V-R=0.47\pm0.01$~mag (derived from data by \citealt{solontoi12-sdss} and \citealt{jewitt15-color}, and filter transformations by \citealt{ivezic07-sdss}).  Therefore, comet \BB{} is a nominal LPC in terms of coma color. For comparison, the nucleus color of the comet measured by \citet{bernardinelli21-c2014un271} is $V-R=0.40\pm0.02$~mag (also shown in Fig.~\ref{fig:color}; color transformed following \citealt{abbott21-des-dr2}). The nucleus and coma colors are consistent at the 2$\sigma$ level, with the nucleus being potentially bluer than the coma. This relationship between the colors of the coma and nucleus environments is in agreement with the average colors reported by \citet{jewitt15-color} for active and inactive LPCs.

\subsection{Brightness and Lightcurve}
In our first image of the comet, the spatial distribution is relatively compact, and a total brightness may be estimated.  Within an 8\farcs9 radius aperture (130,000~km), offset from the nucleus to make a tight fit around the visible dust, we measure $r$=18.14$\pm0.07$~mag.  Converting this brightness to an $r$-band absolute magnitude is done by scaling it to \rh=1~au, $\Delta$=1~au, and to a phase angle of 0\degr{} using the Schleicher-Marcus phase function for cometary dust \citep{schleicher11}, yielding $H_r(1,1,0)=4.96$~mag.

\begin{figure*}
  \plotone{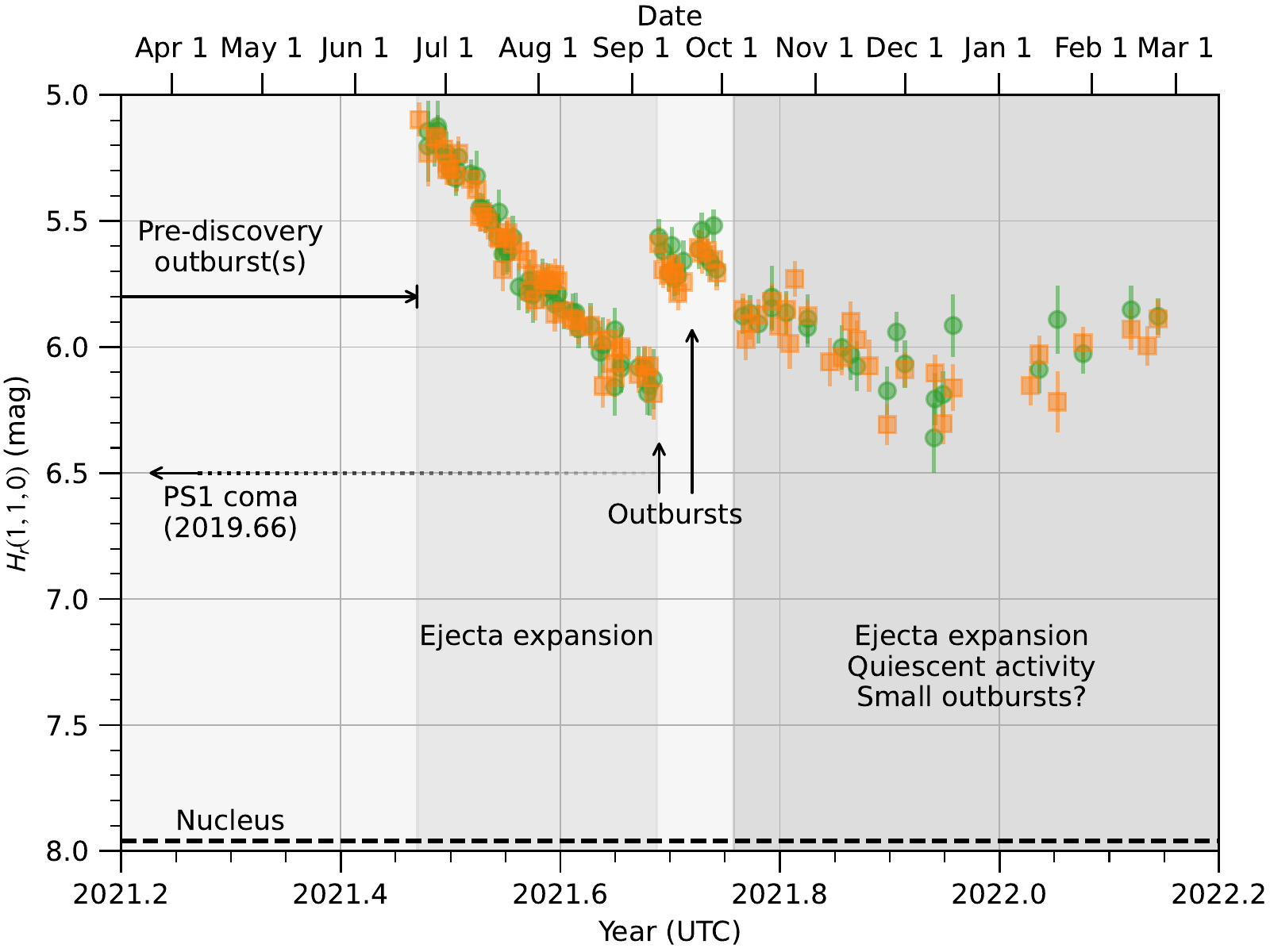}
  \caption{Absolute magnitude, $H_r(1,1,0)$, of comet C/2014 UN$_{271}$ (\BB) measured within 6\arcsec{} radius apertures from LOOK Project photometry ($g$-band as circles, $r$-band as squares) versus time.  Data calibrated to the $g$-band have been scaled with the measured coma colors to make an effective $r$-band data set.  The brightness of the coma from PanSTARRS~1 (PS1) data measured on 2019.66 UTC, using the same aperture size, is marked with a dotted line, and the absolute magnitude of the nucleus is given as a dashed line \citep{bernardinelli21-c2014un271}.}
  \label{fig:lightcurve}
\end{figure*}

In Fig.~\ref{fig:lightcurve}, we present the lightcurve of the comet as absolute magnitude versus time based on our full photometric data set.  Initially, the comet's intrinsic brightness decreased with time, with the absolute magnitude in our photometric aperture increasing from 5.10$\pm$0.07~mag (2021 June 22) to 6.11$\pm$0.04~mag (2021 September 03--08).  This fading period was followed by two brightening events.  The brightening events are likely outbursts, the first occurring between 2021 September~08 02:33 and September~09 22:04 UTC \citep{kelley21-atel14917}, and the second between September 18 02:09 and September 23 04:58 UTC.

With a functional form of the lightcurve, we can characterize the coma fading and the September outburst strengths.  The absolute magnitude within the 6\arcsec{} radius aperture faded at a rate of $13.8 \pm 0.4$~mmag~\inv{day} ($\chi^2_\nu$=0.6, RMS=64~mmag) from 2021 June~22 to September~08.  The September~09 outburst relative strength is $-0.65\pm0.07$~mag and the absolute magnitude of the new material is $H_r=6.32$~mag.  The September~23 outburst strength is derived by comparing pre- and post-outburst $r$-band data, accounting for the linear decline in brightness: $\Delta m = -0.22\pm0.07$~mag, $H_r=7.23$~mag.  Temporal differences of the images indicate that the photometric apertures at the time of the outburst discoveries include all of the new ejecta.  We find no evidence for any other outbursts during this initial fading period.  Each outburst is followed by a decline in coma brightness likely due to expansion of the ejecta and any prior released material beyond the photometric aperture.

Following the outbursts was a period of low-amplitude variability that started in October, with a mean $H_r=5.98\pm0.01$~mag but an RMS of 0.12~mag (19 $r$-band photometric data points between 2021 October 07 and 2022 February 23).  The variability appears to be smooth with time, but gaps in the data could hide small, $\sim-0.1$~mag outbursts.  Monthly averaged absolute magnitudes are $H_r=5.87\pm0.02$, $6.06\pm0.03$, $6.19\pm0.05$, $6.06\pm0.04$, and $5.94\pm0.05$~mag, for 2021 October, November, December, 2022 January, and February, respectively.  Some of the fading from October to November may still be due to residual ejecta from the September outbursts, but without a dynamical model or higher spatial resolution data, or some assessment of possible quiescent activity, this interpretation is speculation.  We will return to this point in Section~\ref{sec:quiescent}.

\subsection{Dust Cross-sectional Area and Mass}\label{sec:mass}
With assumptions on the dust and/or ice grain properties, absolute magnitudes may be converted into total cross-sectional area and mass.  We initially assume all grains have the same properties: effective radius $a_{\mathrm{eff}}$=10~\micron{}, grain mass density of $\rho_g$=1.0~\kgpmmm{}, and $r$-band albedo $A_r$=4\%.  The albedo is a nominal value that is similar to the nucleus surface \citep{Hui2022}, and the grain mass density allows for porous dust grains and/or the presence of water ice.  Thus, the absolute magnitudes of the total coma in June 2021 and the ejecta of the two outbursts in September 2021 ($H_r$=$4.96\pm0.07$, $6.32\pm0.07$, and $7.23\pm0.07$~mag) correspond to cross-sectional areas $G$=$3.1\timesten{5}$, $0.88\timesten{5}$, and $0.38\timesten{5}$~km$^2$, and masses $M$=$4.1\timesten{8}$, $1.2\timesten{8}$, and $0.51\timesten{8}$~kg, respectively (6\% formal uncertainties on all values).  There is considerable uncertainty in the adopted grain parameters, which greatly affects the mass estimates.  The mass scales with the quantity $a_\mathrm{eff}\,\rho_g/A_p$.  Effective grain radius has the largest effect, as sizes of 1 or 100~\micron{}, or even a size distribution, could have been chosen.  We take $a_\mathrm{eff}=10$~\micron{} as our nominal case (equivalent to a differential size distribution $\propto a^{-3.5}$ for $a$ between 1 and 100~\micron, see, e.g., \citealt{ishiguro16-holmes}), and assume a factor of 10 in the mass uncertainty \citep[cf.][]{tubiana15}.

\subsection{Morphological Evolution}

\begin{figure*}
  \plotone{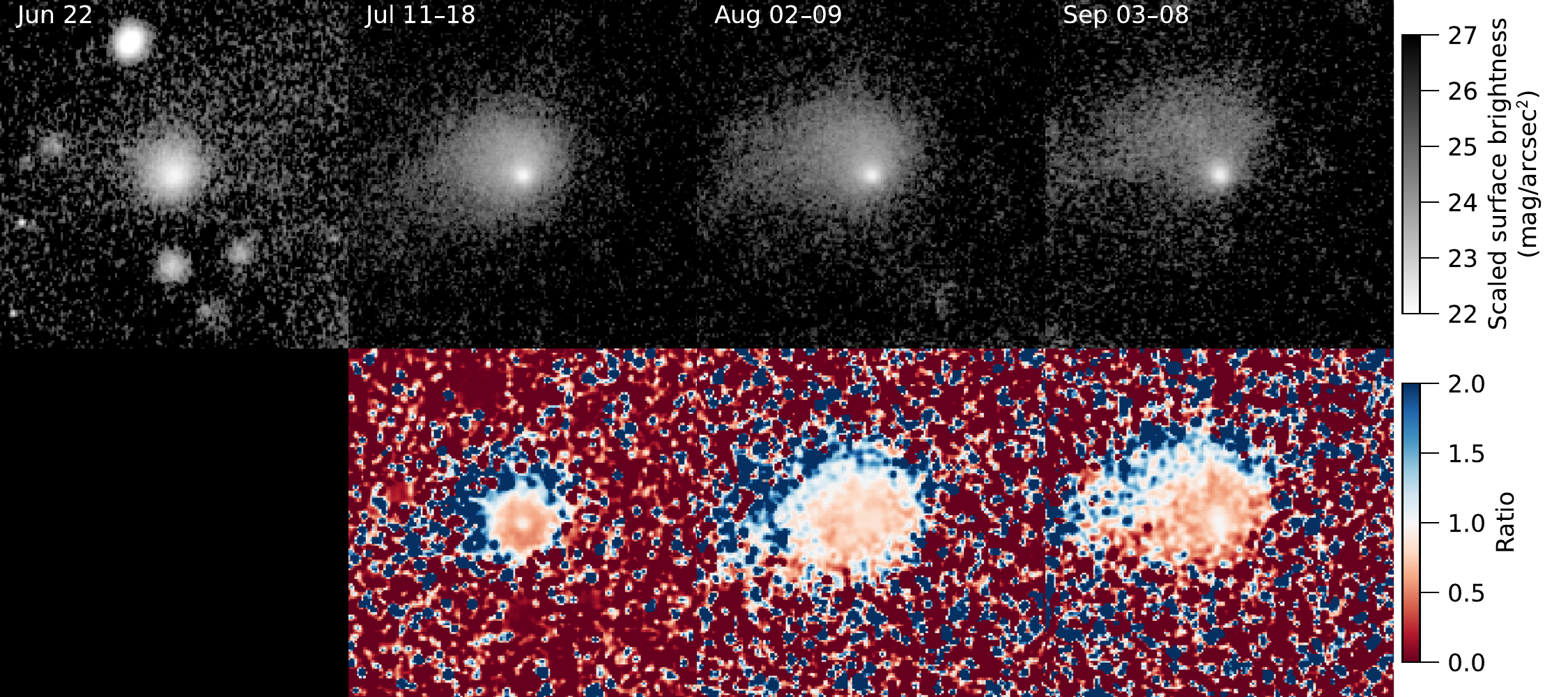}
  \caption{Top: LCO images of comet C/2014 UN$_{271}$ (\BB).  Dates (2021) are as indicated.  Except for the image on June 22, which is a single night $w$-band average, images are median combined $g'$ and $r'$ data taken over a range of dates: 2021 July 11--18, August 02--09, and September 03--08.  All images are displayed with the same photometric color scale, and show an area of 880,000 km $\times$ 880,000 km (60\arcsec{} on 2021 June 22).  Bottom: The same images, but normalized by the previous image in the sequence and smoothed with a $\sigma$=1-pix Gaussian filter.  The expanding morphology of the comet is apparent throughout this period.}
  \label{fig:evol}
\end{figure*}

The comet's morphology changed during the initial fading period (2021 June 22--September 08). To quantitatively evaluate these changes, we median combined $g'$ and $r'$ images taken over the time periods July 11--18, August 03--08, and September 03--08.  Figure~\ref{fig:evol} presents the results.  The time sequence suggests an expanding morphology.  To emphasize the variation, we divided each image by the previous image in the sequence, and show the results in Fig.~\ref{fig:evol}.  The ratio confirms the expansion.

A projected dust expansion speed can be estimated using the co-added images in Fig.~\ref{fig:evol}.  We generated a detection mask on each image, identifying the coma using a 3$\times$3 pixel moving box where at least 1 pixel is more than 2$\sigma$ from the background.  Assuming radial motion, the coma edge expands at a rate of 48 to 84~\mps{} (approximately monotonically increasing from $-$50\degr{} to +100\degr{} east of north) for the time period 2021 June 22 to July 14.  The expansion rate is non-linear with time in our data, falling to near 0 or even negative values thereafter, at least for this edge detection method.  However, this approach does not track any one single feature, but instead the edge of the detected coma, which depends on the image signal-to-noise ratio and the actual surface brightness distribution of the coma.  Furthermore, the expansion may not be radial, and solar radiation pressure or Lorentz forces may be accelerating the dust \citep[e.g.,][]{hui19-c2010u3}.  At best we have estimated the order of magnitude of the initial mean expansion speed, at worst we have measured a lower limit to the fastest moving material.  In addition, an estimate of the slowest material may be made using the lightcurve.  The slowest ejecta takes at least 78~days to move 6\arcsec, thereby moving $\leq$13~\mps{} in projection on the sky.

\begin{figure*}
  \plotone{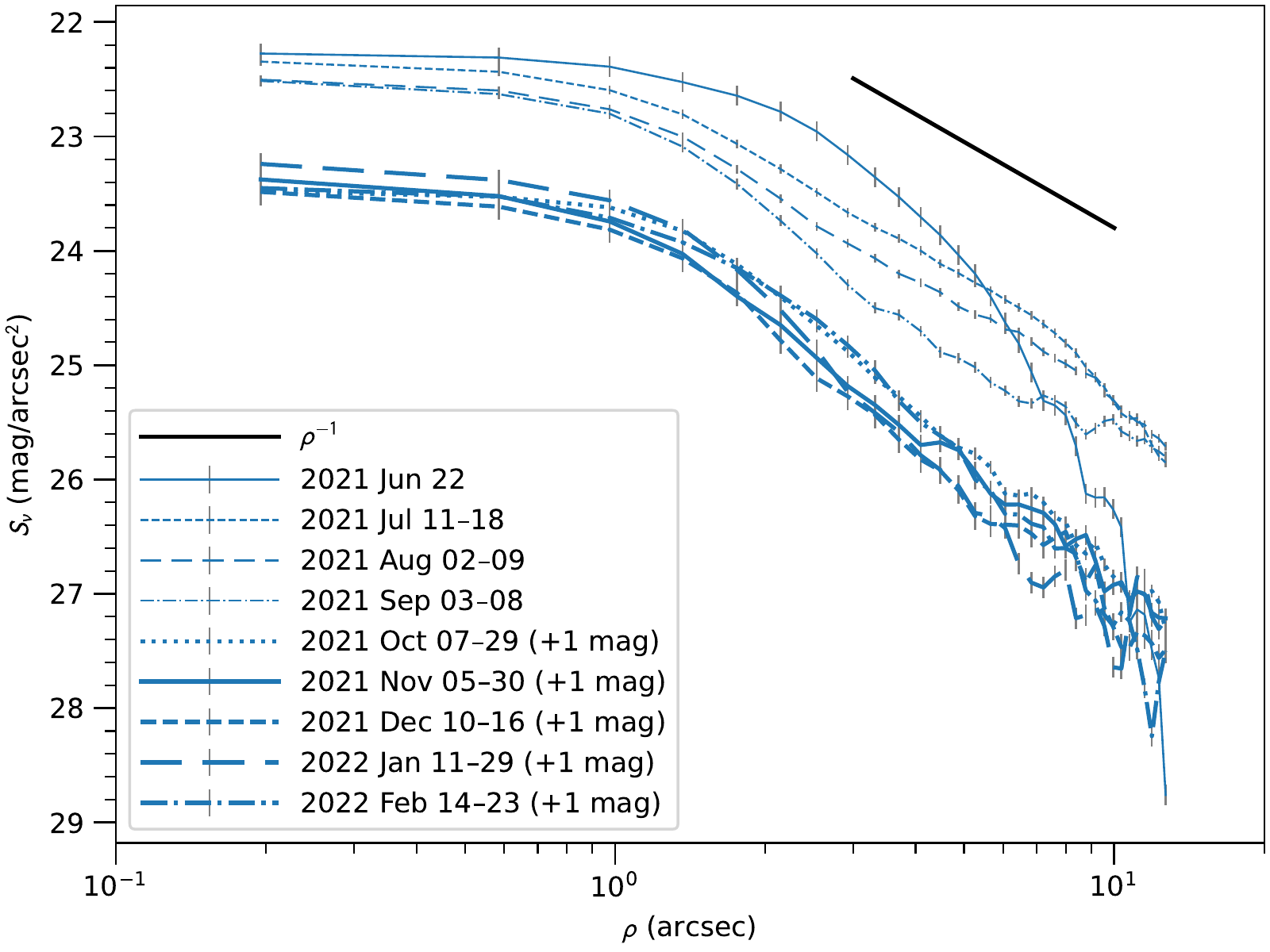}
  \caption{Surface brightness ($S_\nu$) profiles as a function of distance to the nucleus ($\rho$) for the images presented in Fig.~\ref{fig:evol}.  The profile of the comet is initially steep, but flattens with time.  A nominal coma profile, $\rho^{-1}$, is shown for reference.  The mean seeing of these images is 1\farcs8.  Also shown are the monthly mean profiles from 2021 October to 2022 February, offset by $+1$~mag for clarity (mean seeing 2\farcs3).}
  \label{fig:radial}
\end{figure*}

Mean surface brightness ($S_\nu$) profiles as a function of distance to the nucleus ($\rho$) show a morphological evolution of the initial coma (Fig.~\ref{fig:radial}).  The mean profile of the comet flattens with time.  Assuming a functional form of $S_\nu\propto\rho^k$, the best-fit slopes between 3 to 10\arcsec{} are $k=-2.63\pm0.06$, $-1.22\pm0.02$, $-0.99\pm0.02$, and $-0.90\pm0.03$, for the time periods 2021 June 22, July 11--18, August 02--09, and September 03--08, respectively.

Also included in Fig.~\ref{fig:radial} are mean radial profiles for images generated by combining data month-by-month from 2021 October to 2022 February.  At 3 to 10\arcsec, the best-fit radial profile slopes are all steeper than $-1$: --1.39$\pm$0.04, --1.26$\pm$0.06, --1.27$\pm$0.08, --1.71$\pm0.06$ and --1.76$\pm0.06$, for 2021 October, November, December, 2022 January, and February, respectively.  Despite the 50--120~day time period since the last outburst, the coma does not appear to be in a steady-state, and even steepened as the comet brightened in 2022.

\section{Discussion}\label{sec:discussion}
We have characterized the lightcurve and morphology of comet C/2014 UN$_{271}$ (\BB) from 2021 June 22 to 2022 February 23, in an effort to help increase our knowledge of the activity of comets near 20~au.  We now synthesize the results to better understand the state of activity of this particular comet, and compare them to previous work regarding its activity.

\subsection{Outbursts Near 20 au}\label{sec:discuss-outbursts}
Following our first detection of the comet on 2021 June 22, the intrinsic brightness faded, the morphology expanded, and the mean radial profile flattened with time.  Together, these are the signatures of material ejected by a cometary outburst leaving the vicinity of the nucleus.  In contrast, a steady-state coma would have a surface brightness profile proportional to $\rho^{-1}$.  A steep initial slope, $k<-1$, indicates a rapid ejection of new material into the unresolved source.  The fading with time shows that the driver of this new material had ceased, or at least rapidly decreased compared to the timescale of dust expansion.  That the radial profile flattened as the brightness decreased in our photometric aperture is due to the expansion of the material on the sky.  The slope just before the first September outburst, $k=-0.9$, suggests the coma might have been near a steady state at that time, but due to the September outbursts, we could not study the evolution any further to confirm this point (e.g., the profile may have continued to flatten with the expansion of the material).  Therefore, we conclude that the appearance of the comet from 2021 June 22 to September 08 is dominated by an outburst occurring before our first observation.

If these data were instead observations of a previously steady-state coma that had suddenly ceased activity on our about 2021 June 22, then the radial profile would have initially been near $-1$, rather than $<-2$, as observed.  Alternatively, if the comet activity had rapidly increased to establish a new quiescent level by 2021 September, then an initially steep radial profile could have been observed, trending toward --1 with time.  However, in this case the comet would have brightened, not faded with time.

\citet{bernardinelli21-c2014un271} measured the integrated brightness of the comet in PanSTARRS~1 images taken on 2019 August 29 UTC at 22.6~au, 663~days before our first image at 20.2~au.  Their value, $H_r$=6.5~mag in a 6\arcsec{} radius aperture, compares favorably with our minimum brightness in 2021 September, $H_r$=6.1~mag.  Thus, our suggestion that the radial profile at 3--10\arcsec{} in September 2021 was trending toward a steady-state coma is possible.  However, in 2021 November--2022 February, when the comet was at a similar brightness ($H_r$=6.2 to 5.9~mag), the radial profile slope was $-1.3$ to $-1.8$, significantly steeper than $-1$.  We suggest residual ejecta from the September outbursts affect the spatial profile in November--December.  Additional small ($\sim-0.1$~mag) outbursts may have caused the $\sim-0.25$~mag brightening between 2021 December 16 and 2022 February 23.  Validation of this hypothesis would benefit from additional data that could establish the quiescent behavior of the comet.

A comparison of the PS1 photometry from 2019 August to the total coma brightness measured in our first image ($H_r=4.96$~mag) indicates that the outburst preceding our initial observations of this comet was likely no stronger than $\Delta m=-1.5$~mag (6\arcsec{} radius aperture).  Based on the rapid evolution of the ejecta from 2021~June~22 to July 11--18 ($\sim50$~\mps{} expansion speed), the outburst likely occurred in the month before our first observation.  In Section~\ref{sec:mass}, we estimated the total mass of the observed dust on 2021 June 22 to be $4\timesten{8}$~kg, with uncertainties of a factor of 10 due to unknown grain properties.  Removing the PS1 measurement as a possible ambient coma reduces the nominal mass to $3\timesten{8}$~kg, which does not affect the order of magnitude estimate.  In a summary of 17 outbursts over an $\sim8$~year period, \citet{ishiguro16-outbursts} found 3 events (18\%) with cross-sectional areas similar to or larger than our nominal estimate, indicating that this very large cometary nucleus at 20~au has outbursts as large as any typical comet near 1~au.  Outbursts of this scale correspond to equivalent nuclear volumes of cylindrical pits with diameter $D=80(m/10^8~\mathrm{kg})^{1/3}$~m (assuming a bulk nucleus density of 500~\kgpmmm{}, and a depth to diameter ratio of 0.5).  The size of the excavated region is only weakly affected by our large mass uncertainties; a factor of 10 uncertainty in mass corresponds to a factor of 2 uncertainty in diameter. Therefore, such surface features would be comparable to the pits observed on the surfaces of comets 9P/Tempel 1, 67P/Churyumov-Gerasimenko, and 81P/Wild~2 \citep{brownlee04, veverka13, vincent15}.

Without a direct spectroscopic detection of a gas, we cannot determine with certainty what drove the outbursts, or if they were mechanical or thermophysical in nature (i.e., due to thermal cracking, or landslides).  We can likely rule out amorphous to crystalline transitions of water ice as \citet{jewitt17-c2017k2} estimated that they most likely occur within 12.5 au from the Sun for comets (outside of this distance crystallization timescales are too long).  Carbon monoxide ice is the most likely driver of activity and outbursts at 20~au, being that it is both abundant in the cometary population \citep{cochran15} and volatile at this distance \citep{meech04-activity}.  While CO has been observed in distant comets \citep{wierzchos17-echeclus, womack17, yang21-c2017k2}, the relationship between CO and outbursts is not clear, even in the best-observed outbursting comet 29P/Schwassmann-Wachmann 1.  \citet{wierzchos20-sw1} could not identify a strong correlation between their CO and optical lightcurves of comet 29P, and concluded that increases in CO-outgassing are not always involved with the optically identified outbursts.

\subsection{Quiescent Activity and Drivers}\label{sec:quiescent}
Comet \BB{} was clearly active beyond \rh=20~au (Section~\ref{sec:intro}).  Persistent, distant activity of comets is typically attributed to the sublimation of species more volatile than H$_2$O, whose sublimation rates drop off significantly for \rh$>$2.5 au.  \citet{bernardinelli21-c2014un271} estimated the ice responsible for cometary activity, based on photometry of the coma from \rh=29 to 20~au and thermal models of the nuclear surface.  They concluded that the increase of brightness grew rapidly with heliocentric distance, and was consistent with the sublimation of \coo{} ice.  However, their data set included photometry from 2021 June, which we have shown is substantially affected by an outburst that ejected $10^7-10^9$~kg of material.  This outburst brightened the coma within 6\arcsec{} by a factor of 3 to 4, as observed on June 22.  This calls into question the conclusion that \ch{CO2} is the responsible volatile, or at least requires a re-analysis of the activity without using the 2021 data.

The analysis of Hubble Space Telescope data taken 2022 January 08 by \citet{Hui2022} shows that the inner coma ($\rho<1$\arcsec) has logarithmic radial slopes ranging from --1.7 to $\sim$--1.0 and is steepest in the direction of the Sun.  Remarking that a slope of --1.5 is expected in the sunward direction due to radiation pressure, they conclude the observations are consistent with a steady state coma.  In our observations, the absolute magnitude of the comet on 2022 January 11 is consistent with the 2022 December average ($H_r$=6.15$\pm$0.08~mag and 6.19$\pm$0.05~mag, respectively).  Therefore, the brightening seen in 2022 likely occurs after this date and the conclusions of \citet{Hui2022} are not inconsistent with our own.  In fact, it may indicate that our 2021 December photometry is dominated by a quiescent activity state.

\begin{figure*}
  \centering
  \includegraphics[width=0.8\textwidth]{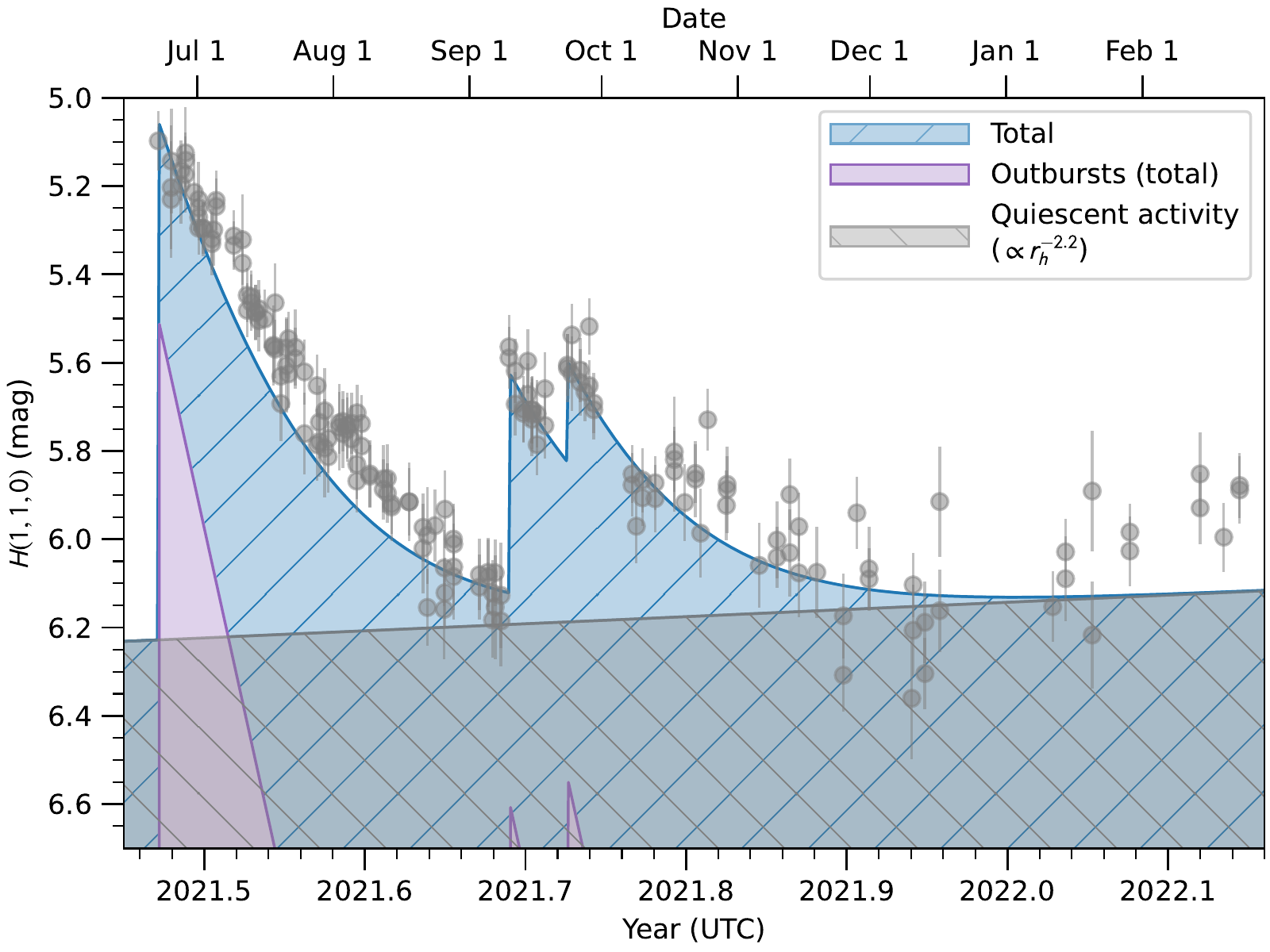}
  \includegraphics[width=0.8\textwidth]{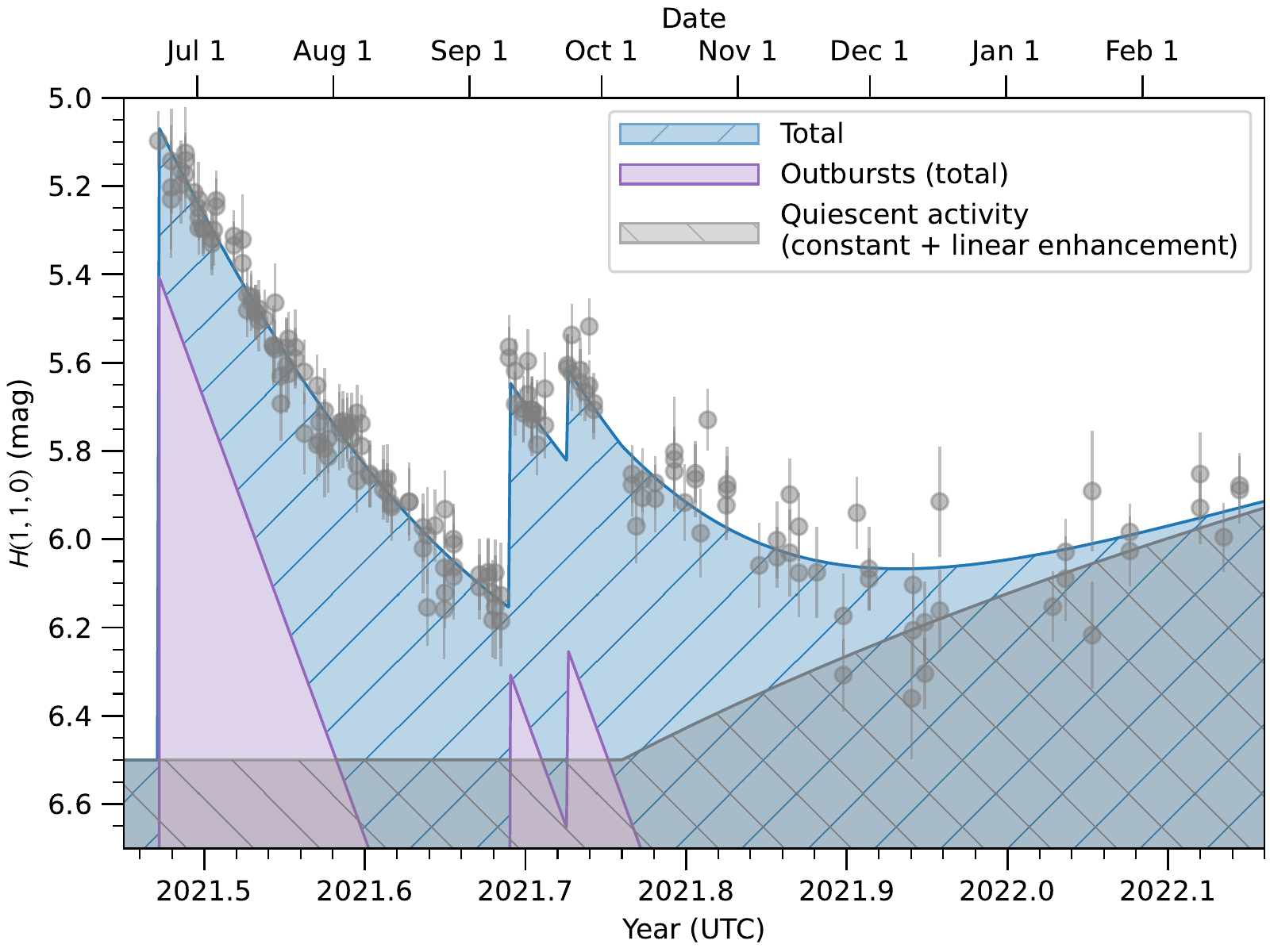}
  \caption{Ad-hoc model lightcurves (absolute magnitude, $H(1,1,0)$ versus time) based on illustrative functional forms for the outbursts and quiescent activity of comet C/2014 UN$_{271}$ (\BB).  LOOK Project photometry of the comet is shown as gray circles ($g$ has been scaled to $r$ as in Fig.~\ref{fig:evol}).  Outbursts follow an exponential decay in dust cross-section, with the same decay timescale, $\tau$, for all events within a panel.  Quiescent activity models are scaled to match the PS1 photometry from 2019 August 28.  Top: Quiescent activity model based on model CO ice sublimation rates, $Z$, parameterized as $Z\propto \rh^{-2.2}$.  Bottom: Quiescent activity model based on constant activity since 2019, and a linearly increasing enhancement starting at 2021.76 (2021 October 5).  Together the models indicate that the lightcurve is outburst dominated but transitions to quiescent activity dominated by 2021 December.}
  \label{fig:lc-models}
\end{figure*}

To assess the contribution of quiescent activity to our data, we have developed an ad-hoc model with illustrative functional forms for each contribution: three outbursts and a quiescent coma.  All three outbursts are considered to be step functions with an exponential fall off in dust cross-section (i.e., magnitude increases linearly with time).  For simplicity, the decay timescales are identical for each event.  For the quiescent activity we take a standard approach in cometary astronomy by assuming the activity scales as a power-law function of heliocentric distance: \rh$^k$.  We derived the power-law exponent from CO ice sublimation rates calculated with the \citet{cowan79} ice sublimation model.  Fits to the model CO sublimation rates between 18 and 26~au yield slopes ranging from $-2.2$ for a slowly rotating nucleus to $-2.4$ for a rapidly rotating nucleus (we assumed a constant active area and a 5\% albedo over all wavelengths; best-fit residuals were $<1$\%).  The dust production rate is assumed to be directly proportional to the CO sublimation rate, and scaled to the PS1 photometry from 2019 August 28.  The outburst strengths and decay timescales are manually adjusted, and the result is shown in Fig.~\ref{fig:lc-models}.

The model lightcurve has poor agreement with the first outburst, either due to our chosen outburst functional form (linear in magnitude space), or to the amount of quiescent activity ($H=6.2$~mag on 2021 September 8).  Here, an outburst decay timescale of $\tau=24$~days is shown.  Longer timescales would better fit the July--August photometry but require a shallower heliocentric distance slope for the quiescent activity in order to fit the early September data points.  However, the quiescent activity is already well below the 2022 January--February photometry ($\Delta H>0.1$~mag), and a shallower slope would cause a greater disagreement. In addition, the assumption that the activity has been growing since the PS1 measurement in 2019 could be incorrect, or that some other effect is needed to match the 2022 data, such as the small outbursts suggested in Section~\ref{sec:discuss-outbursts}.  Furthermore, large aperture photometry from TESS by \citet{bernardinelli21-c2014un271} indicate 1.5$\pm$0.2~mag of brightening from $\rh$=23.8 to 21.2~au (2018 October to 2020 September), more consistent with activity proportional to $\rh^{-12}$ than $\rh^{-2.2}$, or with additional pre-discovery outbursts, which calls into question whether or not the PS1 data at 22.6~au are appropriate for establishing a quiescent activity level.  Altogether, the model suggests that CO ice driven sublimation is possible but some other model assumption must be challenged to confirm this conclusion.

As an alternative to slowly increasing sublimation driven mass loss from CO, we present a second model for the quiescent activity in Fig.~\ref{fig:lc-models}.  Here, we leave the activity as constant since the PS1 measurement, then add a linearly increasing enhancement starting in early October.  This model has a much better agreement with the data.  The outburst decay timescale, $\tau=40$~days, is long enough for the lightcurve to be nearly linear in July--August, consistent with the data.  Moreover, the local minimum in brightness in 2021 December is reproduced, along with the brightening trend in 2022.  We have no particular physical motivation for this model, other than the possibility that the outbursts signaled the onset of a new active area, perhaps due to the propagation of heat into the subsurface as the comet approaches the Sun.

Taken together, our illustrative models in Fig.~\ref{fig:lc-models} indicate that the lightcurve is outburst dominated  from 2021 June through September, and likely quiescent activity dominated by 2021 December.  Observations with the Hubble Space Telescope in 2022 January support this conclusion, and the early assessment that the comet's activity was driven by \coo{} sublimation, which partially relied on outburst-dominated photometry from 2021 June, is unlikely and needs revision.

\section{Conclusions}\label{sec:conclusions}
We examined the intrinsic brightness of comet C/2014 UN$_{271}$ (\BB) as part of a study of this comet's distant activity.  Our data set was comprised of $g'$- and $r'$-band images from the LOOK Project, including data from the Comet Chasers outreach program, over the time period from the discovery of activity on 2021~June~22 to 2022~February~23.  Within 6\arcsec{} radius apertures, the comet had brightened by $\Delta m=$--1.5~mag between the last small-aperture photometric measurements in 2019 August from the PanSTARRS~1 survey \citep{bernardinelli21-c2014un271} and the first LOOK Project photometric data on 2021~June~22.  The absolute magnitude faded with time at a rate of 13.8$\pm$0.4~mmag~\inv{day}, up until two small outbursts on our about 2021 September 09 and 18 ($\Delta m=-0.65$ and $-0.22$~mag).  The two smaller outbursts occurred while material from the June outburst was still expanding and leaving the aperture, i.e the September outbursts happened before the quiescent coma state was reached.  These events were immediately followed by a slight fading in intrinsic brightness ($+0.26$~mag), then a period of low-amplitude variability ($H_r$=5.9 to 6.2~mag), up to the end of our data set.  Radial profiles of the coma suggest the variability may be caused by small, $\sim-0.1$~mag outbursts, although no direct photometric evidence (e.g., a short-scale discontinuity in the lightcurve) is seen.  A comparison of our data to the Hubble Space Telescope observations on 2022 January 08 \citep{Hui2022} suggests our photometry in 2021 December was dominated by the quiescent coma, and an ad-hoc analysis of our lightcurve confirms this point.  Ejecta from the first (prior to 2021 June 22) outburst expanded with speeds of at least 50~\mps{} in the plane of the sky, and the slowest moving ejecta no faster than 13~\mps.  Throughout the observed period, the color of the comet was constant (mean $g-r=0.47\pm0.01$~mag) and consistent with the long-period comet population.

A pre-perihelion outburst ($\Delta m\sim-0.5$~mag) near $\rh \sim20$~au was also observed at comet C/2010 U3 (Boattini) by \citet{hui19-c2010u3}.  It may be that pre-perihelion outbursts are common for Oort cloud comets in the outer Solar System, although none have been reported for the third comet observed at these distances: comet C/2017~K2 \citep{meech17-c2017k2, jewitt17-c2017k2, hui18-c2017k2}.  The nuclei of comets \BB{} and Boattini are likely complex geological worlds, and the cause(s) of these events may be difficult to discern at the present time.  However, further study of cometary activity and outbursts at large heliocentric distances will be beneficial to understanding the long-term surface evolution (e.g., pit formation) of these comets and how it relates to the paucity of Oort cloud comet discoveries \citep[i.e., the long-period comet fading phenomenon;][]{vokrouhlicky19-oort-cloud, kaib22-fading}.  Furthermore, combining future observations with the pre-2021 data, or observations with higher signal-to-noise ratios and finer spatial resolving powers may reveal more about the quiescent activity in this particular comet \citep[e.g.,][]{kokotanekova21-dps, Hui2022}.

\bigskip
We appreciated the helpful comments from the manuscript referee and the scientific editor.  Support for this work was provided by the NASA Solar System Observations program (80NSSC20K0673).  HU, TA, and BW were supported by a UK Science and Technology Facilities Council (STFC) ``Spark Award'' grant for the Comet Chasers school outreach program.

This work makes use of observations from the Las Cumbres Observatory global telescope network as part of the LCO Outbursting Objects Key (LOOK) Project (KEY2020B-009), and the Comet Chasers school outreach program (FTPEPO2014A-004). The Comet Chasers program is part of the Faulkes Telescope Project, which is partly funded by the Dill Faulkes Educational Trust.  Pupils from the following Welsh schools made the observations: St Mary's Catholic Primary School Bridgend, Montgomery Church in Wales School, White Rose Primary School New Tredegar, and Idris Davies School Rhymney.

This research has made use of data and services provided by the International Astronomical Union's Minor Planet Center.

\facilities{LCOGT}

\software{
  astropy \citep{astropy18},
  sbpy \citep{mommert19-sbpy},
  astroquery \citep{ginsburg19-astroquery},
  JPL Horizons \citep{giorgini96},
  SEP \citep{barbary16-sep},
  DS9 \citep{joye03-ds9},
  reproject \citep{robitaille20},
  photutils \citep{bradley21-photutils1.1.0},
  calviacat \citep{kelley19-calviacat},
  ccdproc \citep{craig21-ccdproc2.2.0},
  BANZAI \citep{mccully18-banzai},
  fastrot \citep{vanselous21-ice}
}

\bibliography{apj-jour,references,new-references}
\bibliographystyle{aasjournal}

\end{document}